# Shear of the vector potential in electric-dipole radiation

Keith J. Kasunic

*University of California at Irvine, Irvine, CA 92697*

**Abstract**

Electric-dipole radiation has the well-known far-field solutions of transverse electric and magnetic (TEM) fields with peak values propagating normal to – and going to zero along – the dipole axis. The calculation of these electromagnetic fields is based on the use of the magnetic vector potential: Its time derivative and divergence for the electric field **E**, and its curl for the magnetic field **B**.    In this paper, we suggest the possibility of additional radiation fields, namely, those based on the shear of the vector potential. Our results show that these shear waves have a $1/r$ dependence on far-field distance, with four radiative components: (1)-(3) an orthogonal triad consisting of two transverse waves and a longitudinal wave propagating *along* the dipole axis (analogous to an acoustic dipole); and (4) a transverse wave propagating into the same solid angle as **E** and **B**, parallel to the **E**-field but with one-half the magnitude of the **B**-field. If confirmed by experiment, possible applications include RF communications, novel biomedical sensors, and a new tool in the search for extra-terrestrial intelligence (SETI).

## I. Background

Traditionally, dipole radiation is thought of as a vector-field problem to be solved for **E** and **B**. Radiation from an oscillating electric dipole then has the well-known far-field solutions of transverse electric and magnetic (TEM) fields with peak values propagating normal to – and going to zero along – the dipole axis. The calculation of these electromagnetic fields is based on the use of the magnetic vector potential **A**: Its time derivative and divergence for the electric field **E**, and its curl (or "rotation") for the magnetic field **B**.

We suggest that a tensor solution using the vector potential is more fundamental, and this leads to the possibility of yet a third type of dipole radiation, namely, that of a vector-potential shear wave. Distinct from the Maxwell stress tensor – which is constructed from the **E** and **B**-fields derived from the vector and scalar potentials – we propose that a more complete approach which does not throw away any information is to start with the vector-potential gradient [Grad(**A**)] tensor itself, and decompose it into its irreducible components. Using the results of Thorne and Blandford [1], this decomposition of Grad(**A**) is illustrated in Fig. 1.

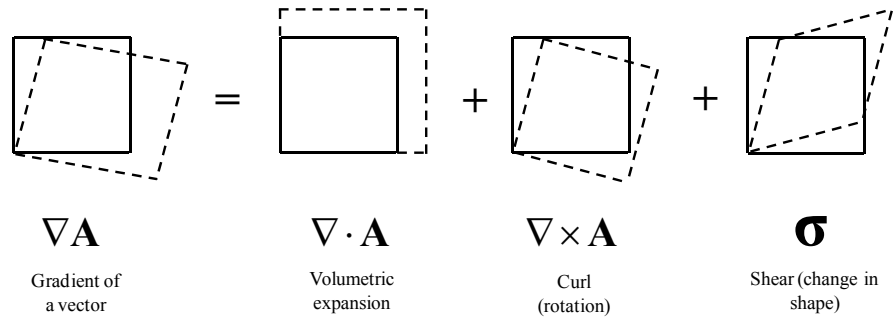

$\nabla \mathbf{A}$ — Gradient of a vector

$\nabla \cdot \mathbf{A}$ — Volumetric expansion

$\nabla \times \mathbf{A}$ — Curl (rotation)

$\boldsymbol{\sigma}$ — Shear (change in shape)

Figure 1 – Vector gradients (second-rank tensors) can be decomposed into volumetric expansion ($\nabla \cdot \mathbf{A}$), rotation ($\nabla \times \mathbf{A}$), and shear ($\boldsymbol{\sigma}$) components. Adapted from Thorne and Blandford [1].

Our hypothesis is thus straightforward: Divergence (a scalar) and curl (a vector) of the vector potential are, by themselves, insufficient for a complete description of the electrodynamic field. Given that only two of the three irreducible components of the Grad(**A**) second-rank tensor in Fig. 1 are utilised in obtaining **E** and **B**, we propose that the third component – the shear component – reveals a third type of radiative wave [2].

Accordingly, we propose in this paper the possibility of additional electric-dipole radiation fields based on the shear of the vector potential. While a second-rank tensor analysis would suggest either a hydrodynamic or elastic-wave analogy, what we are proposing is neither, but is instead a geometric interpretation of a vector-potential shear deformation which results in a force on charged particles. We have proposed a similar interpretation for electron motion in the Aharonov-Bohm effect [3].

Our results show that these shear field amplitudes have a 1/*r* dependence on far-field distance, with four radiative components: (1)-(3) an orthogonal triad consisting of two transverse waves and a longitudinal wave propagating *along* the dipole axis (analogous to an acoustic dipole); and (4) a transverse wave propagating into the same solid angle as **E** and **B**, parallel to and in phase with the **E**-field, but with the much smaller magnitude of the **B**-field. These waves are, of course, neither **E** nor **B**, as our results are not a novel technique for generating old-fashioned electromagnetic waves; instead, they rely on an old-fashioned oscillating-dipole technique, proposed here as also generating vector-potential shear waves.

## II. Electric-Dipole Shear-Wave Radiation

As a brief review, we first outline the standard derivation of obtaining the **E**- and **B**-fields from the vector potential. An oscillating dipole moment $\mathbf{p}(t) = p_o\cos(\omega t)\hat{\mathbf{k}}$ with $p_o = qd$ gives a delayed (or "retarded") vector potential $\mathbf{A}(r,\theta,t)$ pointing along the axis of the dipole (Fig. 2), which in spherical coordinates $(r, \theta, \varphi)$ is given by [4-5]

$$A_r(r,\theta,t) = \frac{\mu_o p_o \omega}{4\pi} \frac{\sin(kr-\omega t)}{r}\cos\theta \qquad (1)$$

$$A_\theta(r,\theta,t) = -\frac{\mu_o p_o \omega}{4\pi} \frac{\sin(kr-\omega t)}{r}\sin\theta \qquad (2)$$

for a wavenumber $k = 2\pi/\lambda$ and symmetry about the dipole axis that gives neither an azimuthal component $A_\varphi$ nor any dependence of $A_r$ or $A_\theta$ on the azimuthal angle $\varphi$. Using the Lorenz gauge [Div(**A**) = –($\partial\phi/\partial t$)/$c^2$] on the divergence of **A** in the dipole induction zone, we find the electric potential $\phi$, the electric field $\mathbf{E} = -[\partial\mathbf{A}/\partial t + \text{Grad}(\phi)]$, and the magnetic field $\mathbf{B} = \text{Curl}(\mathbf{A})$. Under the usual assumptions of $r \gg \lambda \gg d$, we obtain orthogonal radiation components $E_\theta$ and $B_\varphi$ both $\sim \sin\theta/r$ in the far field, with a longitudinal component $E_r \sim 1/r^2$ being very small in comparison. These are well-known results, reviewed in more detail in the standard E&M textbooks [4-5].

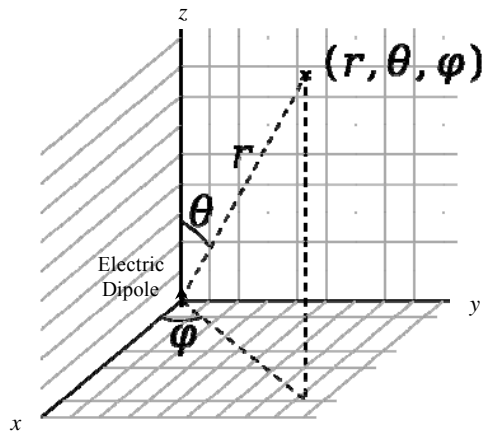

Figure 2 – Geometry of the electric-dipole radiation coordinates $r$, $\theta$, and $\varphi$ with the dipole at the origin and the far-field assumptions of $r \gg \lambda \gg d$ (charge separation distance). Adapted from Wikipedia author Andeggs.

Using the methodology of Thorne and Blandford for gradients of vector fields, we now obtain the radiated shear fields $\sigma_{rr}$, $\sigma_{\theta\theta}$, $\sigma_{\varphi\varphi}$, and $\sigma_{r\theta}$ based on curl-free spatial derivatives of the vector potential. For the radial component normal to the *r*-surface of the wavefront, we use Sec. 11.8 of Thorne and Blandford [1] or the Appendix of Romano and Price [6] to obtain:

$$\sigma_{rr} = \frac{2}{3}\left(\frac{\partial A_r}{\partial r} - \frac{A_r}{r}\right) - \frac{1}{3r}\left(\frac{\partial A_\theta}{\partial \theta} + \frac{A_\theta}{\tan\theta}\right) \quad (3)$$

$$= \frac{\mu_o p_o \omega}{6\pi}\cos\theta\left[\frac{k}{r}\cos(kr-\omega t) - \frac{1}{r^2}\sin(kr-\omega t)\right]$$

showing a longitudinal component of the shear field along the dipole axis. For the θ-component normal to the θ-surface, we find:

$$\sigma_{\theta\theta} = \frac{1}{3}\left(\frac{A_r}{r} - \frac{\partial A_r}{\partial r}\right) + \frac{2}{3r}\left(\frac{\partial A_\theta}{\partial \theta} - \frac{A_\theta}{2\tan\theta}\right) \quad (4)$$

$$= -\frac{\mu_o p_o \omega}{12\pi}\cos\theta\left[\frac{k}{r}\cos(kr-\omega t) - \frac{1}{r^2}\sin(kr-\omega t)\right]$$

For the azimuthal-angle component normal to the φ-surface, we have:

$$\sigma_{\varphi\varphi} = \frac{1}{3}\left(\frac{A_r}{r} - \frac{\partial A_r}{\partial r}\right) + \frac{2}{3r}\left(\frac{A_\theta}{\tan\theta} - \frac{1}{2}\frac{\partial A_\theta}{\partial \theta}\right) \quad (5)$$

$$= -\frac{\mu_o p_o \omega}{12\pi}\cos\theta\left[\frac{k}{r}\cos(kr-\omega t) - \frac{1}{r^2}\sin(kr-\omega t)\right]$$

One common theme illustrated by this orthogonal triad of $\sigma_{rr}$, $\sigma_{\theta\theta}$, and $\sigma_{\varphi\varphi}$ is that they vary in the far field (neglecting $1/r^2$ terms) as $\cos\theta/r$ with respect to the dipole axis ($\theta = 0$). That is, these spherical fields all have a maximum *along* the dipole axis (Fig. 3) – quite distinct from the **E**- and **B**-fields which vary as $\sin\theta/r$, go to zero along the axis, and are a maximum in the equatorial plane perpendicular to the axis ($\theta = 90$ degrees).

Physically, the moving dipole electrons are transferring momentum *m***v** to (and from) the vector-potential field momentum *e***A**, emitting spherical waves along the direction of oscillation. The $\sigma_{rr}$ principal-shear mode (where $\sigma_{r\theta} = \sigma_{\theta r} = 0$) thus acts analogous to an acoustic dipole [7, 8], longitudinally polarised along the dipole axis with an amplitude that varies with $\cos\theta$. We also have two transverse polarisations ($\sigma_{\theta\theta}$ and $\sigma_{\varphi\varphi}$) which have no acoustic analogy.

 

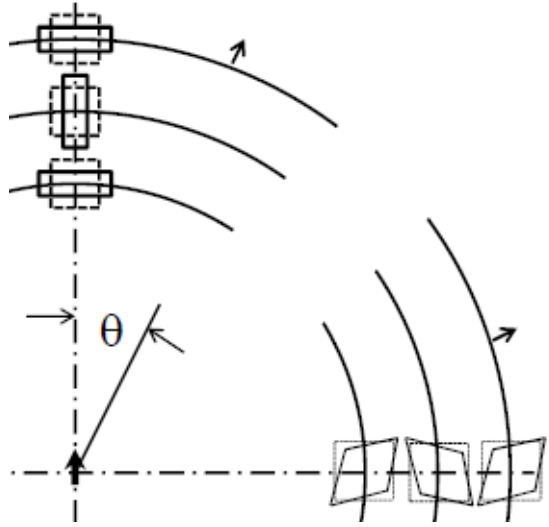

Figure 3 – Shear waves radiating away from an oscillating electric dipole, with the change in shape of a square vector-potential unit cell (dashed lines, not to scale) shown for on-axis ($\theta = 0$) and equatorial ($\theta = 90$ degrees) wavefronts. The unit-cell deformations are due to $\sigma_{rr}$, $\sigma_{\theta\theta}$, and $\sigma_{\varphi\varphi}$ for the on-axis distortion ($\sigma_{\varphi\varphi}$ not shown), and $\sigma_{r\theta} = \sigma_{\theta r}$ for the equatorial.

In addition to these normal components, we also find a tangential component $\sigma_{r\theta}$ on the $r$-surface in the $\theta$-direction (equal to the radial component $\sigma_{\theta r}$ on the $\theta$-surface)

$$\sigma_{r\theta} = \sigma_{\theta r} = \frac{1}{2}\left(\frac{\partial A_\theta}{\partial r} - \frac{A_\theta}{r}\right) + \frac{1}{2r}\frac{\partial A_r}{\partial \theta} \quad (6)$$

$$= -\frac{\mu_o p_o \omega}{8\pi}\sin\theta\left[\frac{k}{r}\cos(kr - \omega t) - \frac{1}{r^2}\sin(kr - \omega t)\right]$$

which varies as $\sin\theta/r$ in the far field i.e., propagates into the same solid angle as the electric and magnetic fields (Fig. 3), but with one-half the magnetic field amplitude $\sim p_o\omega k/r = p_o\omega^2/c_o r$ for $k = \omega/c_o$. This is not surprising, as both the shear and magnetic fields are based on spatial derivatives of the vector potential, while the electric field is based on a time derivative (giving an amplitude $\sim p_o\omega^2/r$).

The *direction* of this shear field, on the other hand, is parallel to – and in phase with – the electric field at all points in space, but an order-of-magnitude smaller than |**E**| by a factor of $c_o$ (the speed of light in vacuum). It is not coupled to the **E**- or **B**-fields in any way; it propagates independently, through the same points in space with (we assume) a phase velocity of $c_o$.

These shear-field tensor components can be organized as a symmetric matrix:

$$\vec{\vec{\sigma}} = \begin{bmatrix} \sigma_{rr} & \sigma_{r\theta} & 0 \\ \sigma_{\theta r} & \sigma_{\theta\theta} & 0 \\ 0 & 0 & \sigma_{\varphi\varphi} \end{bmatrix} \quad (7)$$

where the trace Tr($\sigma$) = $\sigma_{rr}$ + $\sigma_{\theta\theta}$ + $\sigma_{\varphi\varphi}$ = 0, as expected for a symmetric matrix. Physically, the trace represents the divergence of **A** (i.e., volume change of the unit cell), which is zero for the shear tensor [1]. Positive $\sigma_{rr}$ (stretching along the *z*-axis) requires negative $\sigma_{\theta\theta}$ and $\sigma_{\varphi\varphi}$ (compression) to compensate, changing the shape of the unit cell without changing its volume. This is illustrated at the top of Fig. 3 for the on-axis unit-cell deformations. For the equatorial components shown on the right of Fig. 3, the on-axis components are all zero (cos$\theta$ = 0), and the tangential unit-cell deformations along the equator are due to the $\sigma_{r\theta}$ = $\sigma_{\theta r}$ components.

### III. Detecting Shear Waves

Just as with **E** and **B**, it is not **A** that propagates as a wave in dipole radiation. Instead, the propagating wave is created by temporal and/or spatial *variations* of the vector potential: $\partial \mathbf{A}/\partial t$ and Div(**A**) for **E**, Curl(**A**) for **B**, and Shear(**A**) for **σ**. As a result, one approach to detecting shear-wave radiation is to detect these variations, rather than attempting to measure differences in **A** with a Josephson junction via induced phase changes [9].

As with the magnetic field, shear wave amplitudes in Eqns. (3)-(6) are smaller than the electric field by a factor of $c_o$. Given that it is rather difficult to measure such small magnetic fields directly [10-11], how are we to measure the shear field in a possible application such as an RF communications antenna (e. g.)?

Key to measuring the radiated electric field using a conventional RF antenna is the Lorentz force *e***E** on electrons in the antenna; this moves charge through the antenna and creates a measurable voltage difference across a resistive circuit connected to it [12]. Here, we have no Lorentz force for **E** (or **B**), and we need a different detection mechanism than that being used to measure electromagnetic waves.

In a previous paper illustrating the effects of vector-potential shear on electron motion in the Aharonov-Bohm effect [3], we have shown that the force $\mathbf{F}_s$ on an electron due to shear fields depends on the *inner* product of the electron velocity **v** with the shear tensor given by Eq. (7)

$$\mathbf{F}_s = e\mathbf{v} \cdot \vec{\vec{\sigma}} \quad (8)$$

which, distinct from the magnetic field's cross-product force law ($e\mathbf{v} \times \mathbf{B}$), can change the speed of the electron, not just its direction. In addition, the electron velocity gives us an opportunity to compensate for the small value of the shear field with a large velocity. With $|\mathbf{v}| \sim 10^6$ m/sec, e.g., we are only two orders of magnitude reduced from the force due to an electric field.

Applying Eq. (8) to the principal-axis shear waves shown at the top of Fig. 3, we have

$$\mathbf{F}_s = e\left(v_r \sigma_{rr} \hat{\mathbf{r}} + v_\theta \sigma_{\theta\theta} \hat{\boldsymbol{\theta}} + v_\varphi \sigma_{\varphi\varphi} \hat{\boldsymbol{\varphi}}\right) \tag{9}$$

while the force components on an electron from the transverse shear field on the right of Fig. 3

$$\mathbf{F}_s = e\left(v_\theta \sigma_{\theta r} \hat{\mathbf{r}} + v_r \sigma_{r\theta} \hat{\boldsymbol{\theta}}\right) \tag{10}$$

predict a tangential deflection of the electron motion as it moves radially at $v_r$, and a radial deflection as the electron moves tangentially at $v_\theta$.

Electron motion, then, is one approach to directly detecting shear waves. Rather than trying to detect the small forces of Eq. (10) against a large **E**-field background, however, we focus on the normal components of Eq. (9) along the dipole axis where there is neither **E**- nor **B**-field to act as background noise. In principle, reciprocity allows us to point a metal receive (Rx) antenna at the transmit (Tx) dipole antenna and collect a signal. In practice, the electron drift velocities in a metal are rather small, thus making it difficult to extract a useful signal. We propose instead the use of free-space electron emission (Fig. 4), easily allowing electron speeds $\sim 10^6$ m/sec to increase the force in Eq. (9).

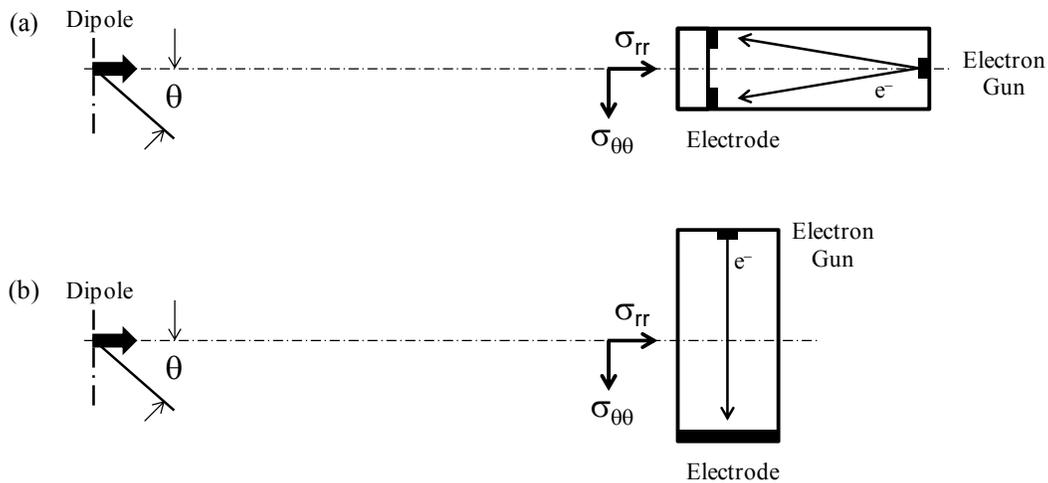

Figure 4 – Two possible architectures for detecting vector-potential shear waves using Eq. (9), measuring $\sigma_{rr}$ in (a) and $\sigma_{\theta\theta}$ in (b). The $\sigma_{\varphi\varphi}$ wave could also be detected using the axis into the page.

Copyright © 2024 by Keith J. Kasunic                                                                                                    Page 7

One possible free-space Rx-antenna configuration detects $\sigma_{rr}$ propagating along the dipole axis [Fig. 4(a)]. The electrons are launched in essentially the $-\hat{\mathbf{r}}$ direction, thus maximizing the signal with a large radial speed $v_r$. A second free-space configuration [Fig. 4(b)] detects one (or both) of the components normal to the dipole axis ($\sigma_{\theta\theta}$ or $\sigma_{\varphi\varphi}$). The figure shows electrons parallel to $\sigma_{\theta\theta}$ for maximum signal; $\sigma_{rr}$ gives no signal, as it is perpendicular to the electron motion. The principal benefit of this architecture is that it allows a larger aperture than Fig. 4(a), thus collecting more energy from the dipole's radiated wavefront.

These architectures also illustrate why a conventional Rx antenna would not detect shear waves. First, a standard broadcast antenna pointing vertically emits most of its electromagnetic power horizontally (~ $\sin\theta$), while the principal-axis shear waves ($\sigma_{rr}$, $\sigma_{\theta\theta}$, and $\sigma_{\varphi\varphi}$) are being emitted vertically (~ $\cos\theta$). Second, even with an airborne antenna – or our eyeballs to sense 0.4-0.7 μm radiation – on an airplane or spacecraft pointing in the correct orientation, the electron speeds required to bring the signal to within a couple of orders of magnitude of the electric field sensed by a conventional antenna are ~ $10^6$ m/sec; any detection of shear waves using slow, drift-velocity electrons in a wire antenna would then be interpreted as background noise.

### IV. Summary and Conclusions

Divergence (a scalar) and curl (a vector) of the vector potential **A** are, by themselves, insufficient for a complete description of the electromagnetic field. We thus view electrodynamics as requiring a more-fundamental tensor description based on gradients in the vector potential [2]. Given that only two of the three irreducible components of the Grad(**A**) second-rank tensor shown in Fig. 1 are utilised in obtaining **E** and **B**, we suggest that the third component – the shear of **A** – reveals a third type of radiative wave emitted by oscillating electric dipoles.

Our results show that these shear waves have a $1/r$ dependence on far-field distance, with four radiative components: (1)-(3) an orthogonal triad consisting of two transverse waves and a longitudinal wave propagating along the dipole axis (analogous to an acoustic dipole); and (4) a transverse wave propagating into the same solid angle as **E** and **B**, parallel to the **E**-field but with one-half the magnitude of the **B**-field.

Just as with **E** and **B**, it is not **A** that propagates from the dipole. Instead, the propagating waves require *variations* in **A**: $\partial \mathbf{A}/\partial t$ and Div(**A**) for **E**, Curl(**A**) for **B**, and Shear(**A**) for σ. The shear fields are completely independent of **E** and **B**, and do not require any modification of Maxwell's equations or the photon concept; any such modifications would require a coupling between the shear and the electromagnetic fields, which we have not addressed in this paper.

Detecting these waves is shown to be possible based on the inner product of electron velocity with the shear tensor. If confirmed by experiment, possible applications include RF communications, novel biomedical sensors, and a new tool in the search for extra-terrestrial intelligence (SETI). As well, there are many questions left unanswered by this brief Letter; we will address many of them – including magnetic dipoles – in future publications.

## Notes and References